\documentstyle[12pt,epsfig]{article}
\newcommand{\qt}{{q}}
\newcommand{\ut}{{u}}
\renewcommand{\kappa}{{k}}
\newcommand{\ba}{\begin{eqnarray}}
\newcommand{\ea}{\end{eqnarray}}
\begin{document}
\begin{titlepage}

\begin{centering}
\begin{flushright}
CERN-TH/2000-234 \\
hep-th/0008074
\end{flushright}

\vspace{0.1in}

{\Large {\bf String-Inspired Higher-Curvature Terms and the Randall--Sundrum
Scenario}}

\vspace{0.1in}

{\bf Nick E. Mavromatos } \\
{\it Department of Physics, Theoretical Physics, King's College London,\\
Strand, London WC2R 2LS, United Kingdom,} \\

\vspace{0.05in}
and
\vspace{0.05in}

{\bf John Rizos} \\
{\it Department of Physics, University of Ioannina, \\
GR 45100 Ioannina, Greece,\\
 and \\
CERN, Theory Division, CH-1211 Geneva 23, Switzerland.} \\

\vspace{0.1in}
 {\bf Abstract}

\end{centering}

\vspace{0.05in}

{\small We consider the ${\cal O}(\alpha ')$ string effective action,
with Gauss--Bonnet curvature-squared
and fourth-order dilaton-derivative terms, which is derived by
a matching procedure with string
amplitudes in five space-time dimensions.
We show that a non-factorizable metric of the Randall--Sundrum (RS) type,
with four-dimensional conformal factor
$e^{-2\kappa|z|}$,
can be a solution of the pertinent equations of motion.
The parameter $\kappa $ is found to be proportional
to the string coupling $g_s$ and thus the solution appears to be
non-perturbative.
It is crucial that the Gauss--Bonnet (GB)
combination has the right (positive in our conventions) sign,
relative to the Einstein term,
which is the case necessitated by
compatibility with string (tree) amplitude computations.
We study the general solution
for the dilaton and metric functions,
and thus construct the appropriate phase-space diagram
in the solution space.
In the case of an anti-de-Sitter bulk,
we demonstrate
that there exists a continuous interpolation between (part of)
the RS solution at $z=+\infty$ and an (integrable) naked singularity
at $z=0$. This
implies the dynamical formation of domain walls
(separated by an infinite distance),
thus restricting the physical bulk space-time to the
positive $z$ axis. Some brief comments on the possibility
of fine-tuning the four-dimensional cosmological constant
to zero are also presented.}

\vspace{0.1in}
\begin{flushleft}
CERN-TH/2000-234 \\
August 2000
\end{flushleft}

\end{titlepage}

\section{Introduction}

Recently considerable effort has been
devoted to the study
of
higher-dimensional space-times with metrics of
non-factorizable form between four-
and higher(bulk)- dimensional coordinates~\cite{others,RS}:
\ba
 ds^2 = e^{-2\sigma(z)} \eta_{ij}dX^i dX^j + dz^2~, i,j=0,1,\dots 3
\ .\label{RSmetric}
\ea
In the modern context of non-perturbative string (brane) theory,
this type of metrics arises from the so-called D(irichlet)-brane
picture of our world,
according to which the observable Universe
is viewed as a three-brane embedded in a higher-dimensional (bulk)
geometry~\cite{dimop,RS}. Among other issues,
in such an approach one looks for mechanisms
that solve the mass hierarchy problem~\cite{RS} or
offer explanations for the vanishing of the (four-dimensional)
cosmological constant. However, the latter case is
inflicted by the presence of naked
singularities in the bulk~\cite{kss} and/or
instabilities~\cite{binetruy}.

In the original approach~\cite{others,RS} the metric (\ref{RSmetric})
has been considered only in connection with Einstein-type theories
of gravitation, i.e. theories in which only the curvature scalar
appears in the gravitational part of the action.
Recently, however, attempts have been made towards the inclusion
of higher-curvature (quadratic) terms in the action~\cite{zee}
of the Gauss--Bonnet type~\cite{zwiebach}.
Such terms,
which arise naturally in (super)string effective actions~\cite{string},
are known to lead to non-trivial cosmological
and general-relativistic solutions, such as singularity-free
expanding~\cite{art} and/or closed~\cite{easter}
universes, and black-hole solutions
with non-trivial (secondary dilaton) hair~\cite{bh,pomazanov}.

In ref. \cite{zee}
five-dimensional
bulk geometries were considered, with our four-dimensional
universe viewed as a three-brane embedded in them.
It was argued, in agreement with the lowest-order (in the scalar curvature)
results~\cite{kss}, that the presence of
higher-curvature Gauss--Bonnet terms
cannot lead to a solution of the cosmological constant on the brane
without fine-tuning, as a result of the appearance of
naked singularities in the bulk.
However,
in the models considered
in ref. \cite{zee},
the Gauss--Bonnet  term in the
action was decoupled from the dilaton
field.
This is {\it not} the case in string-effective models of
higher-derivative gravity, compatible with
string (tree) amplitude computations in the bulk geometry~\cite{string}.
In the latter case, it is known that the dilaton field $\Phi$
couples
to the
higher-curvature part of the effective action
through the appropriate conformal weight, $e^{m\Phi}$.
The weight $m$ is
determined, together with the coefficient of the
GB terms, by the requirement that the effective
action is the one reproduced by the appropriate
string amplitudes~\cite{string}.

The purpose of this work is to reconsider the solution under
the inclusion of proper string-effective
higher-curvature terms.
In this article
we show that, in a set-up where there is an initial
three-dimensional (spatial) brane located at the origin $z=0$
of the bulk dimension of the five-dimensional geometry,
a metric of the form (\ref{RSmetric}) is still a solution
of the equations of motion of an effective action
derived from (conventional) string amplitudes~\cite{string},
up to ${\cal O}(\alpha ')$ in the Regge slope $\alpha '$.
As is well known, such actions can always be cast, by means of
appropriate field redefinitions that leave the (perturbative)
string amplitudes invariant, in a GB form~\cite{zwiebach},
provided one includes appropriate fourth-derivative dilaton terms.
In fact,
as we shall show below, both of  these facts result in
different conclusions, for the
non-constant dilaton case, from those in refs. \cite{zee}.

It is important to note that the sign, as well as the relative strength
$\lambda$
of the GB terms in the action, are uniquely determined
by the amplitude-matching procedure. In our conventions
for the metric and curvature
the coefficient $\lambda$  comes out {\it positive}.
We shall demonstrate that the Randall--Sundrum (RS) type metric~\cite{RS},
with:
\ba
\sigma (z) = \sum _{i} \kappa \left|z - z_i\right|\,,
\label{RSexplicit}
\ea
where $i$ denotes the $i$-th brane, located at $z_i$ along the
bulk direction,
satisfies the equations of motion derived from such an
${\cal O}(\alpha ')$ string-effective action.
It is important to stress that the solution {\it exists only} for
$\lambda >0$. Moreover,
the parameter $\kappa \propto \sqrt{1/\lambda}$.
Since~\cite{string} $\lambda \propto 1/g_s^2$, where $g_s$ is
the string coupling,
the resulting solution appears (formally) non-perturbative.

As we shall show, in our scenario there are also
solutions that are characterized by a {\it vanishing}
vacuum energy contribution {\it on the brane},
a requirement that may come, for instance, by
demanding a consistent embedding of the
solution (\ref{RSmetric}) in a supersymmetric theory
on the brane. However, as argued in
\cite{am}, recoil (quantum) fluctuations
of the $D3$-brane, as a result of scattering
with (bulk) closed strings or other solitonic defects,
may induce supersymmetry obstruction by means
of ``conical'' singularities on the brane.
This yields small contributions to the vacuum energy of the brane,
which, as a result of recoil, finds itself
in an {\it excited} state, rather than
its ground state.  In addition, recoil fluctuations
lead to a dynamical formation of horizons in the bulk
dimension~\cite{leonta} of a given size, which is determined
by the dynamics.
Such
effects, which here are viewed as subleading
to the classical ones we are discussing here, will be the topic
of a forthcoming publication.

In the present article we shall consider dilaton configurations
that depend solely on the bulk dimension $z$.
A particularly interesting case is the one in which the
dilaton
field
is {\it linear} in $z$. This case may be motivated by the fact that
the equations of motion of fields in the geometry (\ref{RSmetric})
acquire a `friction type' form, suggestive of the r\^ole of the
bulk dimension as a renormalization group (RG) parameter~\cite{verlinde},
and actually
of the Liouville-field type~\cite{em,ddk,aben}. The space-like character
of the Liouville field is dictated by the sub-critical
dimensionality of space-time in the specific five-dimensional
geometry under consideration.
Crucial to this interpretation is the fact that the bulk space-time
is of anti-de-Sitter type, which is known to exhibit
holographic properties~\cite{malda}.
The fact that there exist
non-trivial solutions
to the equations of motion, including the (stringy)
Gauss--Bonnet term, is
suggestive of a deeper connection of this string-inspired
approach
with the (holographic) bulk geometries (\ref{RSmetric}).
However, in this paper we shall
not pursue the holographic RG interpretation in detail.
We only mention at this stage that this interpretation does not
seem to hold in the generic case, and requires specific properties
of the bulk geometry (e.g.
the validity of a proper $c$-theorem~\cite{zam,freedman,wittencth}),
which could be quite restrictive in the
presence of higher-curvature terms.
A detailed study of such important issues
will be the topic of a forthcoming publication.

The structure of the article is as follows: in section 2
we formulate the problem, and discuss the GB
higher-curvature combination and its connection
with string amplitudes in a general context of a D-dimensional
space-time. In section 3 we discuss non factorizable
metrics of the form (\ref{RSmetric})
in a five-dimensional set up, with the fifth dimension
generating a bulk geometry, in which a three-brane
world is embedded. In particular, we first demonstrate
the consistency of the Randall--Sundrum-type space-time
with a constant dilaton, in the presence of the Gauss--Bonnet
higher-curvature combination derived from string amplitudes.
We then  proceed to discussing
the linear-dilaton ansatz (with respect to the bulk
coordinate $z$).
We show that the string amplitude
induced Gauss--Bonnet combination {\it is not} consistent
with this solution. However, there is a solution corresponding
to a case~\cite{zee} in which there is {\it no} conformal
coupling of the dilaton with the higher-curvature terms in the
effective action. Our solution, however, still differs from
that of ref. \cite{zee} because of the presence of
fourth-derivative dilaton terms.
Some brief comments on the possibility of fine-tuning the four-dimensional
cosmological constant to zero are made. In particular,
our analysis demonstrates that such fine-tuning is possible
only in the constant dilaton case.
In section 4,
we discuss the general
solution of the equations of motion
for the dilaton and graviton
fields in the string-effective case. This
includes the above solutions
as special cases.
In this general case, one is able of presenting some analytic arguments
on the singularity structure of the solutions, which allow
important conclusions to be drawn
on the underlying physics, that go beyond the
numerical solutions obtained. In particular,
in the string amplitude effective case, we demonstrate
the existence of {\it new solutions} consisting of
continuous functions for the
dilaton and space-time metric fields that {\it interpolate }
between a RS-type solution at $z=+\infty$
and an (integrable) naked singularity at $z=0$.
This implies the {\it dynamical formation}
of domain walls in the bulk geometry obtained
from the string-effective action. The walls  are
separated by an infinite distance, and this results in
a dynamical
restriction of the physical bulk space-time on the positive $z$ axis only.
The fact that this solution emerges from (perturbative)
string-effective actions is remarkable in our opinion, implying that
perturbative world-sheet physics can still lead to
important conclusions of relevance
to (non-perturbative) string theory.
Some conclusions and outlook are presented in section 5.

\section{String amplitude-induced Higher-Curvature Gravity}

In this section we shall formulate the problem mathematically, and set up
our notation and conventions.
Throughout this work we shall follow
the conventions of ref. \cite{dewitt}, according to which
the five-dimensional space-time has signature $(-,+,\dots,+)$, and
the
Riemann tensor is defined as:
${R_{\mu\nu\sigma}}^\tau={\Gamma^\tau_{\nu\sigma,\mu}} - \dots$.

We consider the action:
\ba
S= S_5 + S_4
\label{s5s4}
\ea
where $S_5$ is the five-dimensional part :
\ba
 S_5 &=&
\int d^5x \sqrt{-g} \left[ -R -\frac{4}{3}\left(\nabla_\mu \Phi \right)^2
+ f(\Phi) \left(\alpha R^2 + \beta R_{\mu\nu}^2 +
\gamma R_{\mu\nu\rho\sigma}^2\right) \right.\nonumber \\
&~& + \xi(z) e^{\zeta \Phi}
+ \left. c_2~f(\Phi)\left(\nabla _\mu \Phi \right)^4 + \dots
\right]\ ,
\label{actionGB}
\ea
with $\Phi$ the dilaton field, and
the $\dots$ denoting other types of contraction
of the four-derivative dilaton terms; these will not be of
interest to us here, for reasons that will be explained below.

The four-dimensional part $S_4$ of the action (\ref{s5s4})
is defined as:
\ba
 S_4 = \sum_{i} \int d^4x \sqrt{-g_{(4)}} e^{\omega \Phi} v(z_i)
\label{s4}
\ea
where
\ba
g_{(4)}^{\mu\nu}=\left\{
\begin{array}{l}
g^{\mu\nu}\ , \,\mu,\nu<5\\
0 \ \ \ ,\ \mbox{otherwise}\\
\end{array}\right.
\ea
and the sum over $i$ extends over D-brane walls located at $z=z_i$
along the fifth dimension~\footnote{It is also possible to consider~\cite{leonta}
a `stuck' of such
$D$-branes, in which case $\sum_{i}$ is replaced by $\int dz$
over flat integration measure, and $v(z_i) \rightarrow v(z)$.
This term is not varied with respect to the fifth dimensional (bulk)
gravitational field.}.

The quantities $\alpha,\beta,\gamma, c_2$ are constants
to be determined below by matching with
string amplitudes in the bulk geometry.
We
notice that in our approach we consider the vacuum energy
in the bulk and on the brane as having a specific (exponential)
dependence
on the dilaton field $\Phi$, dictated by string amplitude
computations. More general models, in which one considers
arbitrary scalar potential functions of $\Phi$ have  also been
considered in the literature~\cite{dilatonpot}, but will not be
analysed here. We simply mention that
the precise dynamics behind models with dilaton potentials
is still unknown; in tree-level critical string theory there are no
such potentials, but
string-loop corrections
may be responsible for their generation.

We now consider, for definiteness,
the case in which the action $S_5$ is derived from a ${\cal O}(\alpha ')$
($\alpha '$ the Regge slope) heterotic-type
string theory in the low-energy limit in $D(=5)$-space-time
dimensions.
Some remarks are in order at this point.
From a formal point of view, one may
think of the (bulk) fifth dimension in
the space-time (\ref{RSmetric}) as a (space-like) Liouville
mode~\cite{verlinde,em}.
A more conventional (and probably safer) approach, which we shall
adopt here,
is to assume initially a ten-dimensional space-time, in which
three branes are embedded. In the bulk one may, then,
consider the propagation
of {\it closed} strings only~\cite{dimop}, but take the case in
which
all but one of the
bulk coordinates are compactified. In that case, the induced string theory
amplitudes will formally correspond to those living in
an effective 5-dimensional
space-time, in the sense that one may consider
string backgrounds that depend only on the uncompactified coordinates,
and restrict oneself to
effective string amplitudes (or, equivalently,
$\sigma$-model conformal-invariance conditions~\cite{string})
for those degrees of freedom.

With the above in mind, we have~\cite{string}:
\ba
\alpha =+1, ~~f(\Phi)=\lambda
~e^{\theta\Phi}~,~
\lambda =\alpha '/8g_s^2 > 0\ ,
\label{lambdastring}
\ea
where $g_s$ is the string coupling. In this case we also have
$\zeta=-\theta=\frac{4}{\sqrt{3(D-2)}}$~($=4/3$ in
$D=5$-dimensions of (formal) interest to us here).
Moreover, in (perturbative)
string theory one has the freedom~\cite{string}
to redefine the graviton and dilaton fields so as to ensure that
the quadratic-curvature terms in (\ref{actionGB}) are of the
ghost-free GB form~\cite{zwiebach}:
\begin{equation}
{\cal R}^2_{GB}=R_{\mu\nu\rho\sigma}R^{\mu\nu\rho\sigma} -
4~R_{\mu\nu}R^{\mu\nu} + R^2\ .
\label{GBeq}
\end{equation}
This field-redefinition ambiguity also allows us to consider the
four-derivative dilaton terms in (\ref{actionGB}) as having the single
structure exhibited above. Matching with tree-level string amplitudes
to ${\cal O}(\alpha ')$ then
requires~\cite{string}
\ba
c_2 = \frac{16}{9}\frac{D-4}{D-2}.
\label{tseyt}
\ea
It is interesting to note that for four
dimensional targets, this coefficient vanishes(!).
This fourth-derivative dilaton term will turn out to yield,
in the five-dimensional case, the essential
difference in the solutions obtained here from those in ref. \cite{zee}.

The graviton equations of motion derived from (\ref{actionGB})
in the effective string case are (with $\alpha=\gamma=1, \beta=-4, c_2=16/27$)~:
\ba
&~& 0= R^{\mu\nu} + \frac{1}{2}g^{\mu\nu}
\left(-R - \frac{4}{3}(\nabla\Phi)^2+c_2 f(\Phi) (\nabla\Phi)^4
+ \xi(z) e^{\zeta \Phi}\right) \nonumber \\
&~&+\frac{1}{2}\sum_i{\frac{\sqrt{-g_{(4)}}}{\sqrt{-g}}}g_{(4)}^{\mu\nu}
e^{\omega \Phi} v(z_i)- \frac{4}{3}(\nabla^\mu \Phi) (\nabla^\nu \Phi)
  \nonumber \\
&~&-f(\Phi) \left(2\alpha R R^{\mu\nu} + 2\beta {R^\mu}_\sigma R^{\nu\sigma}
+ 2\gamma R^\mu_{\sigma\tau\rho}R^{\nu\sigma\tau\rho} \right)
\nonumber \\
&~& +\frac{1}{2} g^{\mu\nu} f(\Phi) \left(\alpha R^2 +
\beta R_{\sigma\tau}R^{\sigma\tau} + \gamma
R_{\sigma\tau\rho\kappa}R^{\sigma\tau\rho\kappa} \right) \nonumber \\
&~& + 2\alpha \{ {(g^{\mu\nu} f(\Phi)R)_{;\sigma}}^\sigma -
(f(\Phi)R)_;^{\mu\nu} \}  \nonumber \\
&~& +\beta \{
(g^{\mu\nu}f(\Phi)R^{\sigma\tau})_{;\sigma\tau} +
{(f(\Phi) R^{\mu\nu})_{;\sigma}}^\sigma
- {{(f(\Phi) R^{\mu\sigma})_;}^\nu}_{\sigma}
- {{(f(\Phi) R^{\nu\sigma})_;}^\mu}_{\sigma}\}  \nonumber \\
&~& + 2\gamma \{ (f(\Phi) R^{\mu\sigma\nu\tau})_{;\sigma\tau} +
(f(\Phi) R^{\mu\sigma\nu\tau})_{;\tau\sigma} \} \nonumber\\
&~&-  2c_2 f(\Phi) (\nabla^\mu\Phi) (\nabla^\nu \Phi)
{(\nabla \Phi)}^2
\label{gravitoneq}
\ea
where $;$ denotes covariant differentiation.

The dilaton equation of motion, on the other hand,
yields:
\ba
&~&0= \frac{8}{3}\nabla^2 \Phi + f'(\Phi)
\left(\alpha R^2
+ \beta R_{\mu\nu}R^{\mu\nu} + \gamma R_{\mu\nu\rho\sigma}R^{\mu\nu\rho\sigma} \right)\nonumber \\
&~&+ \sum_i{\frac{\sqrt{-g_{(4)}}}{\sqrt{-g}}}\omega e^{\omega \Phi} v(z_i)
 -
 4 c_2  \nabla_\mu \left(f(\Phi) (\nabla^\mu \Phi) (\nabla \Phi)^2 \right)
 \nonumber \\
&~&+ \zeta \xi(z) e^{\zeta\Phi}+ c_2 f'(\Phi) \left(\nabla \Phi\right)^4\ ,
\label{dilatoneq}
\ea
where the prime denotes differentiation with respect to $\Phi$.

In the next two sections we shall study the classical
solutions of these equations in the context of
non-factorizable space-times of the form (\ref{RSmetric}).

\section{String-Induced Higher-Curvature Gravity and Non-Factorizable Metrics}

\subsection{General Remarks}

We consider the non-factorizable
ansatz (\ref{RSmetric})
for the five-dimensional metric~\cite{others,RS},
which recently attracted a great deal of attention because of
its connection with the  view our world as a D(irichlet)-brane
embedded in the five-dimensional geometry~\cite{dimop,RS}.
Our point in this article is to examine first
whether such metrics are compatible
with the low-energy effective action obtained from
the ${\cal O}(\alpha ')$ string effective action (\ref{actionGB}).
As we shall show below,
it is only for a particular (positive) sign of
the GB term (\ref{GBeq}) relative to the Einstein term,
which is the case obtained from string amplitudes~\cite{string},
that the equations of motion in the space-time
(\ref{RSmetric})
have a real solution. Moreover, we shall also verify
that the specific Randall--Sundrum scenario (\ref{RSexplicit})
is a solution of the equations of motion under certain
conditions.

Assuming
that the metric
function $\sigma (z)$
in (\ref{RSmetric}) and the dilaton fields $\Phi (z)$
are functions only of $z\,$, we write the equations of motion
(\ref{gravitoneq}),(\ref{dilatoneq})
in the form:
\ba
0&=&\frac{e^{\omega \,\Phi (z)}\,v(z)}
   {2} + \frac{e^
      {\zeta \,\Phi (z)}\,\xi (z)}
     {2} - 6\,{\sigma '(z)}^2 +
  12\,e^{\theta \,\Phi (z)}\,
   \lambda \,{\sigma '(z)}^4 \nonumber\\
   &~&-
  36\,e^{\theta \,\Phi (z)}\,
   \theta \,\lambda \,
   {\sigma '(z)}^3\,\Phi '(z) -
  \frac{2\,{\Phi '(z)}^2}{3} +
  12\,e^{\theta \,\Phi (z)}\,
   {\theta }^2\,\lambda \,
   {\sigma '(z)}^2\,{\Phi '(z)}^2 \nonumber\\
&~&+
   \frac{8\,e^{\theta \,\Phi (z)}\,
     \lambda \,{\Phi '(z)}^4}{27} +
   3\,\sigma ''(z) -
  12\,e^{\theta \,\Phi (z)}\,
   \lambda \,{\sigma '(z)}^2\,
   \sigma ''(z) \nonumber\\
&~&+
  24\,e^{\theta \,\Phi (z)}\,
   \theta \,\lambda \,\sigma '(z)\,
   \Phi '(z)\,\sigma ''(z) +
  12\,e^{\theta \,\Phi (z)}\,
   \theta \,\lambda \,
   {\sigma '(z)}^2\,\Phi ''(z)\label{ldeq2}
\ea
\ba
0&=&\frac{e^{\zeta \,\Phi (z)}\,
     \xi (z)}{2} -
  6\,{\sigma '(z)}^2 +
  12\,e^{\theta \,\Phi (z)}\,
   \lambda \,{\sigma '(z)}^4 -
  48\,e^{\theta \,\Phi (z)}\,
   \theta \,\lambda \,
   {\sigma '(z)}^3\,\Phi '(z) \nonumber\\
&~&+
  \frac{2\,{\Phi '(z)}^2}{3} -
  \frac{8\,e^{\theta \,\Phi (z)}\,
     \lambda \,{\Phi '(z)}^4}{9}\label{ldeq5}
\ea
\ba
0&=&e^{\omega \,\Phi (z)}\,\omega \,
   v(z) + e^{\zeta \,\Phi (z)}\,
   \zeta \,\xi (z) +
  120\,e^{\theta \,\Phi (z)}\,
   \theta \,\lambda \,
   {\sigma '(z)}^4+\frac{8\,\Phi ''(z)}{3} \nonumber\\
&~&-
  \frac{32\,\sigma '(z)\,\Phi '(z)}
   {3} + \frac{256\,
     e^{\theta \,\Phi (z)}\,
     \lambda \,\sigma '(z)\,
     {\Phi '(z)}^3}{27} -
  \frac{16\,e^{\theta \,\Phi (z)}\,
     \theta \,\lambda \,
     {\Phi '(z)}^4}{9} \nonumber\\
&~&-
  96\,e^{\theta \,\Phi (z)}\,
   \theta \,\lambda \,
   {\sigma '(z)}^2\,\sigma ''(z)  -
  \frac{64\,e^{\theta \,\Phi (z)}\,
     \lambda \,{\Phi '(z)}^2\,
     \Phi ''(z)}{9}\label{ldeqphi}\ .
\ea
Owing to the Bianchi identities, only two
of the equations are linearly independent
in the bulk.
It is straightforward to verify  the
following relation among the equations:
\ba
&~&
8\,\sigma '(z)\times\left[
(\ref{ldeq2}) -(\ref{ldeq5})\right]-\Phi '(z)\times(\ref{ldeqphi})
+2\,\frac{d}{dz}(\ref{ldeq5})\nonumber\\
&~&\ \ \ \  \ \ \ \ \ \ \ \ \ \ \ \ \ \ \ = e^{\zeta \,\Phi (z)}\,
     \xi '(z) +
    \,e^{\omega \,\Phi (z)}\,v(z)\,
     \left( 4\,\sigma '(z) -
       \omega \,\Phi '(z) \right)\label{redunt}\ .
\ea
Note that,
in order to avoid breaking of Poincar\'e
invariance in the bulk space-time, which we
assume here~\cite{others,RS},
we must impose:
\ba
\xi '(z)=0\ .
\ea
It should also be noted,
however, that it is possible to preserve Poincar\'e invariance
in the bulk by including a more general dilaton potential
$\xi (\Phi)$~\cite{zee,dilatonpot}.
In the (heterotic)string-inspired context, of interest to us here,
such potentials may be generated by string-loop corrections.
We shall not
discuss this case explicitly here, as it will not affect
our qualitative conclusions.

\subsection{Constant Dilaton Case and the Randall--Sundrum Space-Time}

We commence our analysis
with
the case of constant dilaton. In this case,
we can set $\Phi = \eta = {\rm constant}$,
and $\Phi'=\Phi''=0$ in the
bulk, but {\em not on the brane}, since $\Phi'$
can be discontinuous there, as we shall discuss later on.
In this case the equations of motion are reduced to
\ba
e^{\eta \omega} v \delta(z) &=& \frac{d}{dz}
\left( -6\,{\sigma '(z)} + 8\,e^{\eta \,\theta }\,
     \lambda \,{\sigma '(z)}^3 \right)\label{meq}
\ea
\ba
e^{\zeta \,\eta }\,\xi (z) -
  12\,{\sigma '(z)}^2 +
  24\,e^{\eta \,\theta }\,\lambda \,{\sigma '(z)}^4=0\ ,
\label{cosmobulk}
\ea
implying:
\ba
\left( -6\,{\sigma '(z)} + 8\,e^{\eta \,\theta }\,
     \lambda \,{\sigma '(z)}^3 \right)= c = {\rm constant}
\ea
in the bulk.
As a third-degree equation this has always a real solution
of the form: $\sigma '(z)= k_+\,,z>0$, $\sigma '(z)=k_{-}\,, z<0$.

We now integrate (\ref{meq}) over $z$ to an interval that includes
the brane at $z=0$:
\ba
e^{\eta \omega} v = \left.\left( -6\,{\sigma '(z)} + 8\,e^{\eta \,\theta }\,
     \lambda \,{\sigma '(z)}^3 \right)\right|_{0_-}^{0^+} \label{ineq}\ ,
\ea
which reduces to
\ba
e^{\eta \omega} v =  -6(k_+-k_-) + 8\,e^{\eta \,\theta }\,
     \lambda \,(k_+^3-k_-^3)\ ,
\ea
thus relating $k_+,k_-$ with $v$.

Solving (\ref{cosmobulk}) with respect to $\xi$,
by requiring
continuity  of $\xi(z)$ at $z=0$, we obtain
\ba
e^{\eta \zeta} \xi =-12\,k_+^2\,
  \left( -1 + 2\,e^{\eta \,\theta }\,
     \lambda \,k_+^2 \right)=
-12\,k_-^2\,
  \left( -1 + 2\,e^{\eta \,\theta }\,
     \lambda \,k_-^2 \right)\label{rs1}\ ,
\ea
which
has two solutions. The first one is:
\ba
k_+=-k_-=k\ ,
\label{RS2} \ea
which is the RS solution~\cite{RS}~\footnote{Note that
the solution with $k_+=k_-$ has a continuous metric function
at $z=0$, and hence is {\it not} of the RS type.}.

The second solution is:
\ba
k_+^2+k_-^2=\frac{e^{-\eta\theta}}{2\,\lambda}
\ea
and exists only for $\lambda>0$, which is the case
compatible with string amplitude computations~\cite{string}.

From the dilaton equation (\ref{ldeqphi}) in the bulk,
after taking into account (\ref{cosmobulk}),
one has:
\ba
\zeta  + 2\,e^{\eta \,\theta }\,
   \left( -\zeta  + 5\,\theta
     \right) \,\lambda \,{{k_+}}^2=
\zeta  + 2\,e^{\eta \,\theta }\,
   \left( -\zeta  + 5\,\theta
     \right) \,\lambda \,{{k_-}}^2=0\ ,
\ea
which, in conjunction with (\ref{rs1}), leads to {\it either}
\ba
k_+=-k_-=k\ \mbox{and} \ \zeta  + 2\,e^{\eta \,\theta }\,
   \left( -\zeta  + 5\,\theta
     \right) \,\lambda \,{{k}}^2 =0\label{rs2a}
\ea
{\it or}
\ba
\zeta=\theta=0\ \mbox{and}\ k_+^2+k_-^2=\frac{e^{-\eta\theta}}{2\,\lambda}\ .
\label{rs2b}
\ea
Finally, integrating the dilaton equation (\ref{ldeqphi}) in the
neighbourhood of the brane, we obtain
\ba
e^{\eta\omega}\,\omega\,v=
\left.32\,e^{\eta\theta}\,\theta\,\lambda\sigma '(z)^3\right|_{0_-}^{0+}=
32\,e^{\eta\theta}\,\theta\,\lambda\left(k_+^3-k_-^3\right)\ .
\label{rs3}
\ea
From (\ref{rs1}), (\ref{rs2a}), (\ref{rs3}) we thus have:
\ba
 e^{\zeta\eta}{\xi }&=&
     {12\,k^2 - 24\,k^4\,\lambda }\\
    e^{\omega\eta}v&=&
     4\,k\,\left( -3 +
         4\,k^2\,\lambda  \right) \\
   \zeta &=&
     \frac{10\,k^2\,\theta \,\lambda }
       {-1 + 2\,k^2\,\lambda }\\
   {\omega }&=&
     \frac{16\,k^2\,\theta \,\lambda }
       {-3 + 4\,k^2\,\lambda }
\ea
Note that the string solution $\zeta=-\theta~(=\frac{4}{3}~$
 for $5$-dimensional string theory$)$
 is satisfied
for
\ba
\omega=\frac{2}{3}\ ,\ v=-\frac{32\,k}{3}\ ,\ \xi=10k^2\ , \
k=\frac{1}{2\sqrt{3\lambda}} \ , \ \lambda = \frac{1}{8g_s^2}\ .
\label{sol}
\ea
Since $k$ is positive, we observe that the string-effective action
yields, in the case of a constant dilaton, the RS scenario~\cite{RS}
in which the bulk spacetime is of
anti-de-Sitter (in the sense of a cosmological constant $\xi>0$ in
our conventions), whilst the sign of $v$ is opposite to that of $\xi$
(and hence the brane world at $z=0$ has positive tension).
We also notice that the solution (\ref{sol}) implies that
the sign of the conformal weight $\omega$ is opposite to that of
$\theta$, which is expected from
generic considerations in string theory~\cite{string}.

For the second solution (\ref{rs2b}),
one obtains, on account of
(\ref{rs1}) and (\ref{rs3}):
\ba
\zeta=\theta=\omega=0
\ea
and
\ba
v&=&-6\,{k_+} + 8\,\lambda \,{{k_+}}^3 -
  2\,{\sqrt{\frac{1}{2\,\lambda } -
       {{k_+}}^2}} -
  8\,\lambda \,{{k_+}}^2\,
   {\sqrt{\frac{1}{2\,\lambda } -
       {{k_+}}^2}}\label{ssolv}\\
\xi&=&
-12\,{{k_+}}^2\,\left( -1 +
    2\,\lambda \,{{k_+}}^2 \right)\\
k_-&=&
-{\sqrt{\frac{1}{2\,\lambda } -
      {{k_+}}^2}}\ .
\label{sol2}
\ea
Above, we have chosen the negative solution for $k_-$ to ensure
finiteness of the metric at $|z|\rightarrow\infty$.
We observe that the bulk spacetime is again
of the anti-de-Sitter type,
for small $\lambda$, where the perturbative string-effective-action
approach is valid, whilst $v$ and $\xi$ come with opposite signs on
the brane at $z=0$.

At this point it is natural to enquire whether
a vanishing cosmological constant on the brane occurs
by an appropriate choice of the (free) parameter
$\kappa_+$ in the solution (\ref{sol2}).
Indeed, in the case of
a single brane (at $z=0$) the four-dimensional cosmological
constant ($\Omega$) is given by:
\ba
  \Omega \equiv \int _{-\infty}^{\infty} \sqrt{-g}  \xi + v =
\frac{k_- - k_+}{2k_-\,k_+}\xi + v\ ,
\ea
which yields the Randall--Sundrum solution, with
\ba
   k_+=-k_- = k =\frac{1}{2\sqrt{\lambda}}
\ea
as the unique solution that guarantees the
zero cosmological constant in our framework,
where higher-curvature corrections have been taken into account.

So far we have concentrated on the case of a single brane,
located at $z=0$.
The above conclusions are not affected by including more than
one branes, as in the approach of \cite{RS}, which
is needed
for a solution of the hierarchy problem.
Within our framework,
despite
the small value of
$\kappa = \sqrt{2/3} g_s$ (\ref{sol}),
in units of $\alpha '$,
this can be achieved by
placing another brane at $z=r_0$,
which we
assume describes the location of the observable world~\cite{RS}.
As in ref. \cite{RS},  $r_0$ may be taken
(within the classical framework) to be
a free parameter, which may be assumed, much
larger than the string scale $\ell _s = \sqrt{\alpha '}$.
In such a case, the mass hierarchy in our world arises from the fact that
the value of the determinant in front of the matter
lagrangian on the brane, at $z=r_0$, will be suppressed
by exponential factors of the generic form $e^{-\kappa r_0}$.
These
can be small for $r_0/\sqrt{\alpha '} $ sufficiently large.

However,
as we shall discuss in the next section,
the general solution to the equations of motion
for the string-effective case, imply the possibility
of a
{\it dynamical } appearance of a second brane
(domain wall) located at
a distance, which is determined by the
underlying dynamics, and in fact turns out to be
{\it infinite}. We should mention that
similar restrictions
on a dynamically-induced magnitude of
$r_0$ may be encountered
in case one consider when
quantum (recoil) fluctuations in the D--branes are considered~\cite{leonta}.
We reserve a discussion
on this  problem
for a forthcoming publication.

\subsection{Linear Dilaton in Randall--Sundrum space-times}

In this subsection we shall examine the simplest possible case
of a non-constant dilaton,
namely that of
a dilaton linear in the bulk dimension~\cite{ddk,aben}:
\ba
\Phi(z)=Q z  + \eta\ , \label{ldil}
\ea
with $Q$ constant.

Considering this case may seem
well-motivated by the proposal on the identification of the
bulk coordinate $z$ as a holographic renormalization group
parameter~\cite{verlinde}, in case
the bulk space-time is anti-de-Sitter, which is known to exhibit
special holographic properties~\cite{malda}.
From this point of view, the linear dilaton ansatz, for a metric
of the form (\ref{RSmetric}), is suggestive of a more specific situation,
namely that of the identification of $z$ with a (space-like)
Liouville mode~\cite{em} in the five-dimensional context.
However, this identification requires some thinking, and is not always
possible. Renormalization-group flow in stringy $\sigma$-model
is irreversible, due to the loss of information in modes
beyond the ultraviolet (world-sheet) cut-off.
This implies the presence of a c-theorem~\cite{zam},
whose existence
for
generic bulk space-times
is not clear at present~\cite{freedman,wittencth,zee}.
We shall not discuss this interpretation
further in this article.
This will be the topic of a forthcoming publication.

Nevertheless in this section we shall
study the linear dilaton case (\ref{ldil}) {\it per se}
and discuss whether this ansatz is compatible
with the metric (\ref{RSmetric}) in the
context of the ${\cal O}(\alpha ')$
string-effective action (\ref{actionGB}).
To this end, we first
consider
the linear combination $(\ref{ldeq2})-3\times(\ref{ldeq5})$
and substitute the ansatz (\ref{ldil}) for the dilaton and the RS
metric (\ref{RSexplicit}).  In such a case, we obtain in the bulk:
\ba
 -72\,Q^2 + 64\,e^{\theta \,\Phi (z)}\,Q^4\,
   \lambda  + 648\,e^{\theta \,\Phi (z)}\,
   k^3\,Q\,\theta \,\lambda  +
  648\,e^{\theta \,\Phi (z)}\,k^2\,Q^2\,
   {\theta }^2\,\lambda
\ea
From this, it is trivial to conclude that the linear dilaton solution
is compatible {\em only} with $\theta=0$, which, on account of the
equations of motion leads to $\zeta=\omega=0$.
In this case, we find two solutions for $\lambda>0\,$ :
\ba
Q^2=\frac{9}{8\,\lambda}\ ,\ k^2=\frac{1}{2\,\lambda}\ ,\
\xi=\frac{1}{4\,\lambda}\ ,\ v=-18\sqrt{\frac{2}{\lambda}}\ ,
\label{sol1a}\ea
or
\ba
Q^2=\frac{9}{8\,\lambda}\ ,\ k^2=\frac{2+\sqrt{6}}{8\,\lambda}\ ,\
\xi=0\ ,\ v=3\,\sqrt{\frac{52+22\sqrt{6}}{\lambda}}
\label{sol1b}\ea
We now remark that,
from the point of view of a possible holographic renor\-malization-group
interpretation of the bulk geometry~\cite{verlinde,malda}, the
consistent solution would be the
first one (\ref{sol1a}), characterized by an anti-de-Sitter
type bulk geometry.
Because of  the $z$-independence of
$Q$ in this case, the fixed points connected
with the renormalization-group flow (i.e. the theories living on
the two branes in the RS geometry) would be degenerate,
being characterized by the
same value of the central charge $Q$, and hence
would be connected by marginal operators in a RG sense on the world-sheet.
This case is common in superstring theories.

Note the relative sign difference in $v$ between the two solutions.
Also notice that in neither of the above cases it is
possible to fine-tune the parameters so as to obtain a
vanishing cosmological constant on the four-dimensional world.
The cosmological constant is relatively small,
for weakly coupled strings, as being proportional
to $g_s$. However, this is not phenomenologically acceptable,
unless one considers (non-realistic) very weakly coupled string theories.

As a final remark, we would like to stress the
difference of our scenario from those discussed in ref. \cite{zee}.
In our case, in contrast to
that discussed in \cite{zee},
there exists the non-trivial fourth-derivative
dilaton term $\left(\nabla \phi \right)^4$. Its presence is crucial
in ensuring (for the $\theta =0$ case) the consistency of the linear-dilaton
ansatz with the non-factorizable metric case, and moreover in
yielding solutions for $\sigma (z)$ that go beyond the
RS scenario.

\section{Beyond the Randall--Sundrum scenario}

In this section we shall examine the general solution, within the
string effective action framework,
for the space-time (\ref{RSmetric}), where we shall treat both
$\sigma (z)$ and the dilaton $\Phi (z)$ as unknown functions,
without restriction to the specific form
of the RS metrics (\ref{RSexplicit}). We shall discuss
the general solution of the
equations of motion (\ref{ldeq2})--(\ref{ldeqphi}),
and discuss the connection with the metrics (\ref{RSexplicit})
as a special case.
As we shall demonstrate
below, sufficient analytic information on the structure
of the solutions can be obtained, which allows us to draw
some general conclusions on the underlying physics.

\subsection{General Solution of the Graviton and Dilaton Equations}

It is convenient  to
use the notation:
\ba
y(z) \equiv \lambda e^{\theta \Phi (z)}~, \quad
u(y) \equiv \frac{1}{\sqrt{y}}\,\frac{d y(z)}{d z}~,\ ~~~ q \equiv\sqrt{y}\,\frac{d\sigma (z)}{d z}~,\, ~~~\Xi=\xi\,\lambda^{-\frac{\zeta}{\theta}}
\label{notations}
\ea
From now on, we shall concentrate on  the case of
string theory $\zeta=-\theta~(=4/3$ in the case of five-dimensional strings).
In the above parametrization, eq. (\ref{ldeq5})
becomes algebraic :
\ba
&~&
16\,\Xi  - 192\,{{\qt}}^2 +
  384\,{{\qt}}^4 -
  1536\,{{\qt}}^3\,
   {\ut} +
  12\,{{\ut}}^2 -
  9\,{{\ut}}^4=0\ .\label{alg}
\ea
Solving (\ref{ldeq2}),(\ref{ldeqphi})
with respect to $\qt'$, $\ut'$ and using (\ref{alg}) we obtain
\ba
y\,\frac{d \qt(y)}{dy}&=&\frac{A(\qt,\ut)}{8\,\ut\,C(\qt,\ut)}\nonumber\\
y\,\frac{d \ut(y)}{dy}&=&\frac{B(\qt,\ut)}{\ut\,C(\qt,\ut)}\ ,
\label{finaleq}
\ea
where
\ba
A(\qt,\ut)&=&
-\left( -4\,{\qt} +
      48\,{{\qt}}^3 -
      2\,{\ut} -
      16\,{{\qt}}^2\,
       {\ut} +
      {{\ut}}^3 \right) \,
    \left( 128\,{{\qt}}^3 -
      2\,{\ut} +
      3\,{{\ut}}^3 \right)
\nonumber\\
B(\qt,\ut)&=&
4\,{\qt}\,
  \left( 1 - 4\,{{\qt}}^2 +
    12\,{\qt}\,
     {\ut} \right) \,
  \left( -4\,{\qt} +
    48\,{{\qt}}^3 -
    2\,{\ut} -
    16\,{{\qt}}^2\,
     {\ut} +
    {{\ut}}^3 \right)
\nonumber\\
C(\qt,\ut)&=&
-2 + 8\,{{\qt}}^2 -
  512\,{{\qt}}^4 -
  16\,{\qt}\,
   {\ut} +
  3\,{{\ut}}^2 -
  12\,{{\qt}}^2\,
   {{\ut}}^2 +
  24\,{\qt}\,
   {{\ut}}^3\ ,
\label{finaleq2}
\ea
Dividing the two equations in (\ref{finaleq}) we obtain:
\ba
\frac{d\qt}{d\ut}=\frac{128\,\qt^3-2\,\ut+3\,\ut^3}{32\,\qt(-1+4\qt^2-12\,\qt\,\ut)}\
.
\ea
Notice that the same equation is obtained by simply differentiating the
algebraic equation (\ref{alg}) with respect to $\ut$, thus demonstrating
that this equation provides the general solution $\qt=\qt(\ut)$. This is a
one-parameter family of solutions, with the parameter being provided
by the bulk cosmological constant $\Xi$.
This result was to be expected, considering, the fact that the three equations are
not independent (c.f. (\ref{redunt})).
Using the result for $\qt(\ut)$ we can formally solve (\ref{finaleq}) for $\ut(y)$
\ba
y= y_0\,{\rm
exp}\left(\int\,d\ut\,\ut\,\frac{C(\qt(\ut),\ut)}{B(\qt(\ut),\ut)}\right)\
,
\ea
from which, on account of (\ref{finaleq}) and (\ref{notations}),
we obtain $y(z)\,(\Phi(z))$ and $\sigma(z)$ as functions
of the bulk coordinate $z$. However, in practice the analysis is obscured
by the presence of divergences in the derivatives of $(q,u)$, which are
resolved only in the physical parametrization ($\sigma(z), \Phi(z)$). In
this case we resort to numerical integration of the full system
(\ref{ldeq2})--(\ref{ldeqphi}).

We first
note that the above equations admit in the bulk some exact
solutions,
which are known analytically.
The first example is the trivial case of
\ba
\Xi~=~\ut=~\qt~=~0~,
\label{gensol0}
\ea
corresponding to
a flat bulk space-time
with a constant dilaton, which is
obviously an exact solution of the equations of motion.

A second exact solution occurs for anti-de-Sitter
bulk with a specific value of the cosmological constant:
\ba
\qt^2=\frac{1}{12},\ ~\ut=0,\ ~\Xi=\frac{5}{6}\ .
\label{RSsol}\ea
This is the Randall--Sundrum (constant
dilaton) solution (\ref{sol}), derived in subsection 3.2.

A third exact solution can be found by inspecting eqs
(\ref{finaleq}),(\ref{finaleq2}). We notice that both $dq/dy$ and
$du/dy$ vanish for
\ba
q=q_0,\ u=u_0\ne0\ {\rm with}\ -4\,{\qt_0} +
    48\,{{\qt_0}}^3 -
    2\,{\ut_0} -
    16\,{{\qt_0}}^2\,
     {\ut_0} +
    {{\ut_0}}^3 =0
\ea
and thus the above points correspond to exact solutions with $\Xi$ determined
from (\ref{alg}). These solutions
that corresponds to a curve in the phase space (see discussion in subsection 4.3  and
figure \ref{rssfig}).
In terms of the metric and
the dilaton these solutions are singular
\ba
\sigma(z)=\sigma_0+\sigma_1\,{\rm ln}(z-z_0), \
\Phi(z)=-\frac{3}{4}{\rm ln}\phi_0-\frac{3}{2}\,{\rm ln}(z-z_0), \
\label{exact3}
\ea
with $\phi_0={1 + {{\sigma }_1}}/
  ({2 - 8\,{{{\sigma }_1}}^2 +
    12\,{{{\sigma }_1}}^3})$
and $\Xi$ determined form (\ref{alg}) in the range
$0.60<\Xi<44.44$.

Finally, another exact solution is
\ba
\qt=0, ~~~\ut=\pm\sqrt{2},~~~ \Xi=\frac{3}{4}
\ea
or, in terms of the original parameters,
\ba
\sigma (z)={\rm const},\ ~~ \Phi (z) =-\frac{3}{2}~{\rm ln}(\frac{z}{\sqrt{2\,\lambda}}),
~~\xi(z)=\frac{3}{4\,\lambda} \label{gensol2}
\ea
which implies a flat bulk space-time, with a non-constant dilaton.

\begin{figure}[htb]
\begin{center}
\epsfig{figure=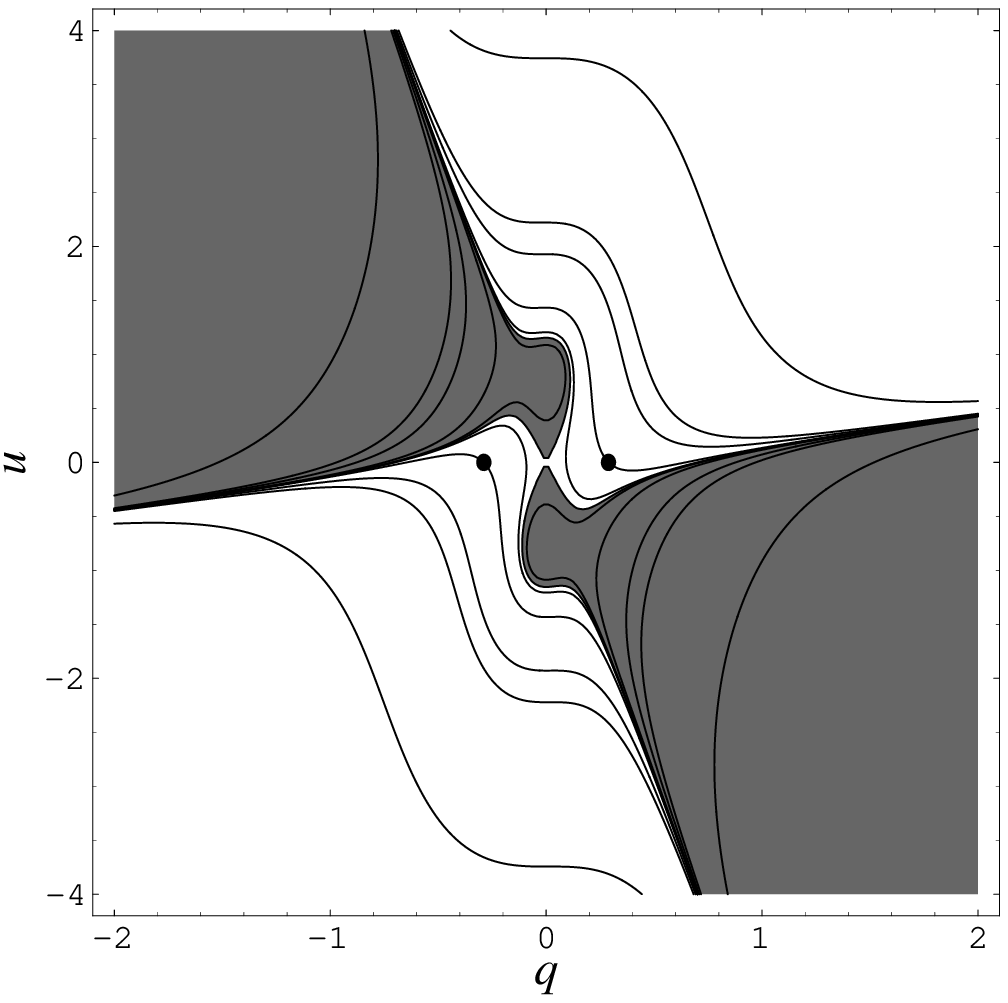}
\caption{\it Phase-space diagram
$\ut(\qt)$ for the general solution
of the five-dimensional equations of motion for dilaton and
graviton fields in the presence of
${\cal O}(\alpha ')$ terms in the string-inspired
effective action. The various contours are parametrized
by the values of the bulk cosmological constant ($\Xi$).
The shaded
region corresponds to de-Sitter bulk space-times ($\Xi < 0$),
while the rest of the
diagram corresponds to anti-de-Sitter bulk ($\Xi >0$).
The boundaries between these two regions correspond to the
$\Xi=0$ contours.
The dots represent the Randall--Sundrum spacetime.
The origin ($u=q=0$) corresponds to a flat bulk space-time.}
\label{ps}
\end{center}
\end{figure}

The general solution of
 (\ref{finaleq}) and (\ref{alg})
is represented by a (Escher-illusion-like
\footnote{See M.C.~Escher, ``Liberation'', lithograph (1955),
for instance at {\tt http://www.\\WorldOfEscher.com/gallery/Liberation.html} } ) phase-space diagram given in
figure {\ref{ps}}.
The shaded region correspond to de-Sitter type bulk, $\Xi< 0$,
whilst the rest of the graph correspond to the case of interest here,
namely anti-de-Sitter type bulk, $\Xi > 0$.
The various contours in the
diagram of fig. \ref{ps}
correspond to solutions with various values of the
cosmological constant $\Xi=\lambda\xi$. For instance,
the depicted contours in the anti-de-Sitter region
in the upper-right-side of the graph correspond
to the following values of $\Xi$:0, 0.1, 5/6, 5, 10, 100,
in increasing sequence, pointing outwards from the
center of the graph.

The above-discussed exact solutions (\ref{gensol0})-(\ref{gensol2})
correspond to specific points
in that diagram. For instance,
the trivial flat space-time appears at the origin $(0,0)$
of the
solution space, whilst
the RS solution (\ref{RSsol}) corresponds to the marked points
in figure \ref{ps}.

From the graph it becomes clear that there are {\it only}
four singular points,
corresponding to the cases $\qt \to \pm \infty,
\ut \to \pm$, in which $\qt /\ut \sim {\rm const} $.
We shall study these points analytically in the next subsection.

\subsection{Singularity structure in ($\ut,\qt$) parameter space}

In this section we perform an analytic study
of the singular points of the
solution space of (\ref{finaleq})
in the $(\ut,\qt)$ parametrization.
It would be instructive to consider first the
connection between singularities in the physical parameters
$\sigma, \Phi $ and their derivatives and the transformed space
parameters $q, u, y$ and their derivatives.
One easily concludes from (\ref{notations})
that
when $q, u$ diverge  there are divergences in at least one of
the quantities $\Phi, {d\sigma}/{dz}, {d\Phi}/{dz}\,$ \footnote{
The inverse is also true, that is finite $q$ and $u$ correspond
to finite $\sigma, \sigma'$ and $\Phi, \Phi'$ with exception the points of the
exact solution (\ref{exact3}).}.

Regarding the second derivatives we have
\ba
\frac{d^2\sigma}{dz^2}=u\left(\frac{dq}{dy}-\frac{q}{2y}\right),
\
\frac{d^2y}{dz^2}=u\left(\frac{du}{dy}+\frac{u}{2y}\right),
\ea
and thus singularities in $dq/dy, du/dy$ for $u,q,y=$finite,
correspond to singularities in the derivatives of the physical
parameters in all cases except $u\to0$ with $u (dq/dy)=$finite and
$u (du/dy)=$finite.

We are now ready to proceed to a study of the singularities.
The right-hand-side of equations (\ref{finaleq}) does not contain any
explicit dependence on $y$ and, thus, can be easily examined for
singularities at the cases $\qt\to\pm\infty$ and/or
$\ut\to\pm\infty$ and/or $\qt\to0$ and/or $\ut\to0$. After a
systematic search (see fig. \ref{ps})
of all cases we find only one class of four singular
solutions at $\qt\sim\ut\to\pm\infty$
\ba
\qt\sim\rho_i\ut,\ ~~~ \ut\sim u_0 y^{c_i}\ ,
\ea
where $\rho_i, i=1,2$ is one of the real solutions of the equation
$-3 - 512\,\rho^3 + 128\,\rho^4=0$, $\rho_1\approx-0.178$, $\rho_2\approx 4.000$
and
\ba
c_i=\frac{1 - 16\,\rho_i^2 + 48\,\rho_i^3}
  {32\,\left( -3 + \rho_i \right) \,\rho_i^2}\ ,
  \ea
$c_1\approx-0.070$, $c_2\approx5.500$.

We thus conclude
(c.f. (\ref{notations})) that the singularities in the
solution space are
encountered at $y\to0$ $(\Phi\to+\infty$)
for $\rho_1$ and
$y\to+\infty$  $(\Phi\to-\infty$) for $\rho_2$.
From (\ref{notations}) one observes
that, near the singular points,
the dilaton $y$ ($\Phi (z)$) and metric $\sigma (z)$
functions acquire the form:
\ba
&~&y^{\frac{1}{2}-c_i}=\left(\frac{1}{2}-c_i\right)u_0z > 0, ~~~
y(z)=\left(u_0 (\frac{1}{2}-c_i) z\right)^{\frac{1}{\frac{1}{2}-c_i}}~,
 \nonumber \\
&~& \Phi (z) =
\Phi_0 -\frac{3}{2-4c_i}{\rm ln}|z|~,
\nonumber \\
&~&\sigma (z) = \sigma_0 + \frac{\rho_i}{\frac{1}{2}-c_i}~{\rm ln}|z|
\label{metricdil}
\ea
where $\sigma_0,u_0,\Phi_0 ={\rm const.}$
Several remarks are in order at this point~:~(i) {\it Both} singularities
occur at $z=0$, but we have the condition $u_0z >0$  (for $\rho_1$),
and $u_0z<0$  (for $\rho_2$), so if one assumes a fixed sign of $u_0$
for both singularities, then we see that
one approaches $z=0$ from different side for each type of
singular solution.
(ii) As one approaches the singularities,
both the dilaton and metric functions
diverge logarithmically with $z$,
(iii) The scalar curvature in the five-dimensional bulk near the singularity
is given by:
\ba
R=4\left(5(\sigma '(z))^2 - 2 \sigma ''(z)\right)=
4(5\gamma - 2)\gamma~\frac{1}{z^2}~,
~~\gamma =\frac{\rho_i}{\frac{1}{2}-c_i}\ .
\label{curvscal}
\ea
We observe from (\ref{curvscal}) that the curvature
diverges as $z \to 0$, and thus one has a {\it naked} singularity
there. However, for a four-dimensional observer, living on the brane
at $z\to 0$, the singularity for both cases
is {\it integrable},
given that the (covariant) integral of the scalar curvature
over $z$ in the vicinity of the singularity yields
\ba
    \int dz \sqrt{g} R (z) \propto \int _{z \sim \epsilon \to 0} dz z^{-4(\gamma_i +1/2)} \sim
\epsilon^{-4\gamma_i -1} \to 0\ ,
\ea
since the exponent $-4\gamma_i -1$ takes on the positive values $0.25$ and $2.2$ for the $\rho_1$ and $\rho_2$ cases respectively.

\begin{figure}[htb]
\begin{center}
\epsfig{figure=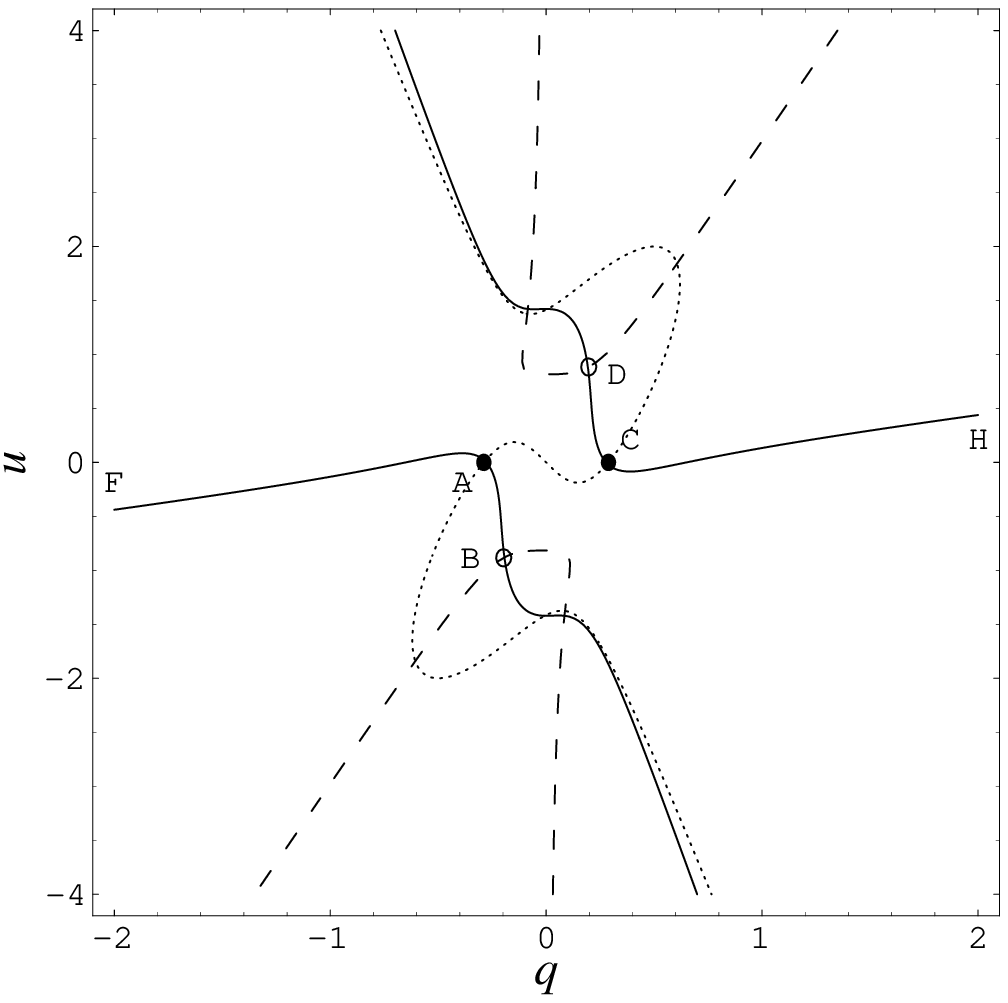}
\caption{\it A solution interpolating between the Randall--Sundrum
solution at $z=+\infty$
and a naked singularity (see (\ref{metricdil}) for the case $\rho_2$)
at $z=0$. There are two (physically equivalent) branches
CH and AF.
There are
other solutions that
encounter (D or B) non-resolvable  singularities in the derivatives
$dq/dy, du/dy$ (dashed curves). The dotted curve (for $u\ne0$) corresponds to the
exact solutions of eq. (\ref{exact3}).}
\label{rssfig}
\end{center}
\end{figure}

By closely inspecting the general solution  we observe
(see figure \ref{ps} and for a detailed view fig.\ref{rssfig})
 that the RS solution
(\ref{RSsol}) is not isolated, but is connected
by means of a continuous interpolating function
with the naked singularity $\rho_2$ (\ref{metricdil}).
The important issue is to determine the point in the
axis $z$  to which the RS solution corresponds.
This will be the topic of the next subsection.

Before doing so we should remark that
there are other branches of the solution space that connect
the RS solution with genuine singularities of the derivatives
of $\qt$ and $\ut$ at points in which the $C(q,u)$ factor in the
denominator of (\ref{finaleq}) vanishes, for $\qt,\ut$ finite
and non zero. This becomes evident from figure \ref{rssfig} where
we plot the contours crossing the RS points $(A,C)$ as well as
the curves (dashed lines) representing the above (non-resolvable) divergences
in the derivatives $dq/dy, du/dy$.
Such points may correspond to naked singularities in the
$\sigma (z)$ and $\Phi (z)$ space, given that the
first and second derivatives of both fields diverge,
assuming finite values of $y$.

We have also plotted in figure \ref{rssfig} the curve that
represents the (one parameter) exact solution of equation (\ref{exact3})
(dotted curve). The various points on the curve correspond to exact
solutions with different values of $\Xi$. The curve does not contain
the points $A, B$, since the $u=0$ points are excluded.

In this article we shall not discuss  these
branches of the general solution further. A detailed study
is postponed to a future publication.

\subsection{Interpolating between
the Randall--Sundrum solution and Naked Singularities}

Let us now proceed to an analytic
determination of the behaviour of the solution
in the neighbourhood of the RS points ($A, C$
in fig. \ref{rssfig}), which will also
determine the point in the $z$ axis to which
the RS solution corresponds.

Expanding (\ref{alg}) around $q=q_0=\pm\frac{1}{\sqrt{12}}$, $u=0$ we
obtain $q\sim \qt_0 -\frac{1}{2} u$, which on account of
(\ref{finaleq}),(\ref{finaleq2})
leads to:
\ba
\qt &=& \pm\frac{1}{2\sqrt{3}} \mp \frac{1}{\sqrt{3}}{\rm ln}\frac{y}{y_0}~,
\nonumber \\
\ut &=& \pm \frac{2}{\sqrt{3}}{\rm ln}\frac{y}{y_0}
\label{qusol}
\ea
with $y\to y_0=$finite.
The sign of $y-y_0$ determines the branch of the solutions
depicted in figure \ref{rssfig}.
For the $AF$ or $CH$ branches, which we shall study here,
$y~<~y_0$, on account of (\ref{qusol}) (the opposite is true
for the $AB$, $CD$ branches).

From (\ref{notations})
we have:
\ba
     y=y_0 - ~e^{\pm\frac{2}{\sqrt{3}y_0}z}~,
\label{ydef}
\ea
which implies
that the point $y\to y_0$ occurs
at $z \to \mp\infty$,  for $y_0$
finite and $\qt_0=\pm\sqrt{1/12}$ respectively.

Solving for the dilaton $\Phi(z)$ and metric $\sigma (z)$ functions,
we then obtain:
\ba
&~& \Phi (z) \simeq \Phi _0 +
\frac{3}{4 y_0}~e^{\pm\frac{2}{\sqrt{3~y_0}}~z }
+ \dots~, \nonumber \\
&~&\sigma (z) = \sigma_0 \pm \frac{1}{\sqrt{12 y_0}}~z +
\frac{1}{8 y_0}~e^{\pm\frac{2}{\sqrt{3~y_0}}~z} + \dots~,
\label{rss}
\ea
where $\Phi_0 \equiv
-\frac{3}{4}{\rm ln}\left(\frac{y_0}{\lambda}\right)$.

Thus, we see that the leading parts of the solution (\ref{rss}),
for infinite $z$ (e.g  $z \to +\infty$ for $q_0=-1/\sqrt{12}$),
is a smooth Randall--Sundrum type (with $k=-1/\sqrt{12 y_0}$), which
should be understood only as the part of the RS metric (\ref{RSexplicit})
inside a given region of the bulk space-time, bounded by a membrane
located at a position $z\to \infty$
in the bulk. In fact the solution is valid only near the membrane
wall, and deviations from it are exponentially suppressed
with $z$.
The reader should not be alarmed by the
apparent divergent form of the metric element as $z\to \infty$.
The correct way of viewing (\ref{rss})
is to consider first the solution as valid for
$z=\Lambda $, where $\Lambda$ is larger than any other length scale in the
problem. Then, one may shift $z \to {\tilde z}=\Lambda-z$,
and arrange the constant $\sigma_0$ of eq.  (\ref{rss})
to be such as to cancel factors of $\frac{\qt_0}{\sqrt{y_0}}\Lambda$.
Eventually, one may take the limit $\Lambda \to \infty$.
The resulting metric is of the RS type around ${\tilde z}=0$,
whilst the naked singularity now occurs at ${\tilde z}=\infty$.

At this point we should also remark, that inspection of
the phase-space diagram of
figure (\ref{ps}) reveals that the interpolation of the RS solution
passes through the point $\ut=0$ {\it twice}.
In the journey from $z=+\infty$ towards finite values,
the solution passes first through another point $z_0>0$
that has $\ut=0$,
before reaching the
naked singularity $\rho_2$ (\ref{metricdil}) at $z=0$.
In the point $z=z_0$ the behaviour of {\it both} $\Phi(z)$ and
$\sigma (z)$ functions is perfectly {\it regular}.
Indeed, this second point of vanishing $\ut$ occurs
for
$q\to q_1= \pm~\sqrt{5/12}$.
Expanding around this point,
one obtains
$z-z_0={\cal O}\left((y-y_1)^{1/2}\right)$,
for $y\to y_1$,
where $z_0 >0$ is finite.
From (\ref{notations}), then,
it is evident that for $z \sim z_0$:
$\Phi(z) = \Phi_1 - {\cal O}(z^2)$, and
$\sigma(z) = {\rm const} + {\cal O}(z)$,
where $\Phi_1=\frac{1}{\theta}{\rm ln}(y_1/\lambda)$ is a constant.
This is a perfectly regular behaviour in the $z$ space.

\begin{figure}[htb]
\begin{center}
\epsfig{figure=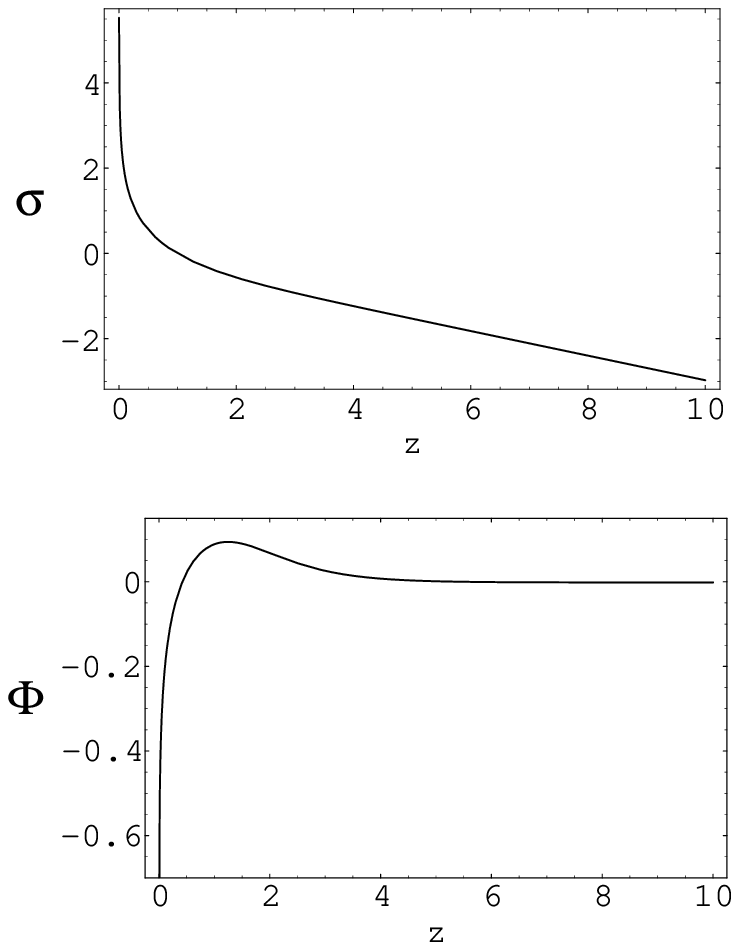}
\caption{\it The metric $\sigma (z)$ and dilaton  $\Phi (z)$ as
functions of $z$
interpolating between a RS type solution ($\qt_0=-\sqrt{1/12}$)
at $z=+\infty$ and a naked singularity (see (\ref{metricdil}) for
 $\rho_2$ )
at $z=0$.
The existence of this solution implies the dynamical
restriction of the bulk space-time to the positive $z$ axis.}
\label{rszfig}
\end{center}
\end{figure}

The
numerical analysis summarized in fig. \ref{rssfig}
indicates that there exist  smooth functions for $\Phi(z)$ and
$\sigma (z)$
interpolating between the RS solution at $z=+\infty$
and the naked (integrable) $\rho_2$ singularity (\ref{metricdil})
at $z=0$. These are plotted in
fig. \ref{rszfig} for the case $q_0=-\sqrt{1/12}$.

The existence of the interpolating solution depicted in figure
\ref{rszfig} implies an important fact about the
nature of the solution in the context of the string-effective
action in which it was derived. The induced bulk space-time
is {\it dynamically} restricted on the positive $z$ axis
(for definiteness, if one considers the $\qt=-\sqrt{1/12}$ branch,
which corresponds to the contour segment
AF in fig. \ref{rssfig}).
In this scenario, our (flat) four dimensional world is viewed as
the boundary of the anti-de-Sitter bulk ($\Xi=5/6$) located
at $z=\infty$. A RS-type solution is valid near our world,
which however deviates from it as $z$ runs
towards the origin $z=0$,  to become
an integrable  naked singularity there.

In this scenario, we observe that the dynamics of the
${\cal O}(\alpha ')$ perturbative string theory
yielded important information on the structure of
the bulk space-time, which may be related with
solitonic (non-perturbative) structures such as D-branes.
The non-perturbative nature of the solution we have thus obtained
becomes clear from the fact that in terms of the original parameters
of the model, the $\kappa$ conformal parameter of the
RS solution (\ref{RSsol}) is found proportional to $1/\sqrt{\lambda} \sim g_s$,
where $g_s$ is the string coupling.

From the point of view of a holographic RG interpretation of the bulk
coordinate ($z$), we remark that the solution of figure \ref{rszfig}
satisfies a ``c-theorem'' in the sense of ref. \cite{freedman}. Namely,
we observe that $\sigma''(z)>0$ for $0<z<\infty$, which implies that
the weakest energy condition is satisfied for this portion of the
bulk space-time.
\section{Conclusions and Outlook}

In the present article we have performed a systematic study of
non-factorizable metrics of the form
(\ref{RSmetric}) in the specific case of
five-dimensional geometries. We have considered the
situation in which such geometries are derived as consistent
solutions of the equations of motion of string effective actions
in the five-dimensional case, to ${\cal O}(\alpha ')$
in the Regge slope. Such terms include
quadratic-curvature contributions
of the Gauss--Bonnet type, as well as fourth-order dilaton derivative terms.

Our analysis has
shown that
it is indeed possible to find compatibility of
such a string-inspired model with
the
Randall--Sundrum scenario, upon the appropriate embedding
of three branes in the five-dimensional space-time.
In addition, we were able to
find more general situations,
which interpolate between the RS metric at the boundary
of an anti-de-Sitter bulk and an (integrable) naked singularity
at the origin.
Such scenarios imply the dynamical formation of domain walls
in the space-time, which may be useful when one discusses
the consistent embedding of D(irichlet)-branes in such a picture
(as is the case of the original RS scenario). In our solutions
the conformal parameter $k$ of the RS type metric, as well
as the bulk cosmological constant, turn out to be proportional
to the string coupling.

There are many issues that remain to be checked.
First the stability of the solution against the inclusion of higher
order $\alpha'^2$ corrections as well as string-loop corrections.
Moreover, in the present work we have assumed that the dilaton
and metric functions
depend only on the bulk coordinate, and we took the four-dimensional
world to be flat. The extension to
more complicated metrics, especially time dependent, is needed
in order to discuss cosmological implications~\cite{art,zee,ovrut}.
Moreover, the proper inclusion of quantum fluctuating (recoiling)
D-branes, in the way discussed in \cite{leonta},
a situation that undoubtedly
is expected to be encountered in a complete quantum theory,
are very interesting issues that
deserve special attention and are
currently under investigation.

In addition,
the precise connection of the bulk coordinate
with a holographic renormalization-group parameter in
the case of anti-de-Sitter bulk geometries also merits
a separate study.
As mentioned in the text, one should re-examine carefully
this interpretation
in the context of the existence of
a proper c-theorem, expressing the
irreversibility of the renormalization group flow in the bulk.
In ref. \cite{freedman}, this c-theorem was suggested to be
provided by the monotonicity of $\sigma'(z)$ ($\sigma'' (z) \ge 0$)
in the metrics (\ref{RSmetric}), which is the result of
a positive energy theorem for consistent matter to be
placed in the bulk. For the interpolating solution of figure \ref{rszfig}
this has been shown to be valid. However, this is not always
true~\cite{zee,wittencth} for generic bulk (anti-de-Sitter) geometries,
especially in the higher-curvature context discussed here,
where the presence of the Gauss--Bonnet terms complicates the
positive energy conditions~\cite{bh}. A detailed study of
such issues will appear in a forthcoming publication.

\section*{Acknowledgements}

The work is partially supported by the European Union
(contract ref. HPRN-CT-2000-00152).
The work of N.E.M. is also partially supported by PPARC (UK).

\end{document}